\documentstyle{elsart}
\begin{document}
\input psfig.sty

\hoffset=0pt
\voffset=0pt
\hyphenation{cal-o-rim-e-ter}
\hyphenation{ped-es-tal}

\begin{frontmatter}
%\centerline{\psfig{figure=draft.eps}}
%\vglue -28cm

%\title{Simple method for determining pedestal noise \\
%in modular independent-channel detectors}
\title{ADC common noise correction and \\
zero suppression in the PIBETA detector}

\author[UVa]{E. Frle\v z\thanksref{author}},
\thanks[author]{Corresponding author; Tel: +1--804--924--6786, 
fax: +1--804--924--4576, e--mail: frlez@virginia. edu (E. Frle\v{z})\hfill}
\author[UVa]{D. Po\v cani\'c},
\author[UVa,PSI]{S. Ritt}
\address[UVa]{Department of Physics, University of Virginia, 
Charlottesville, VA~22904-4714, USA}
\address[PSI]{Paul Scherrer Institut, Villigen PSI, CH-5232, Switzerland}

\begin{abstract}
We describe a simple procedure for reducing ADC common noise in 
modular detectors that does not require additional hardware. 
A method using detector noise groups should work well for modular 
particle detectors such as segmented electromagnetic calorimeters,
plastic scintillator hodoscopes, cathode strip wire chambers,
segmented active targets, and the like. We demonstrate a ``second pedestal 
noise correction'' method by comparing representative ADC pedestal 
spectra for various elements of the PIBETA detector before and after 
the applied correction. 
\par\noindent\hbox{\ }\par\noindent
\par\noindent\hbox{\ }\par\noindent

\end{abstract}
\end{frontmatter}
\vfill\eject

\section{Introduction}
\medskip
Experimental areas in particle accelerator facilities are notoriously noisy
places due to beamline elements like magnets, slits, pumps, power supplies, and
fans as well as other assorted electronics equipment that either runs continuously 
or is intermittently turning on and off. All of the above may cause voltage 
ripples that significantly compromise photomultiplier tube (PMT) 
current pulses digitized by analog-to-digital converters (ADCs). The problem
is compounded especially in analysis of data from segmented detectors where total 
energy and/or direction of each measured particle are derived from smeared ADC values 
of several adjacent detector modules. An experimenter might try to follow the recipes 
for reduction of pedestal noise couplings by designing recommended magnetic shielding 
and applying proper electrical grounding principles~\cite{Ott76,Hor89,Joh90}. 
But such efforts are in practice usually met with only limited success.

In order to minimize the electronic noise arising from ``dirty'' electrical grounds 
and from cross-talk between adjacent detector ADC channels, quality coaxial 
cables are used for connecting PMT anode outputs with inputs of 
fast electronics units. This arrangement though does not protect against 
stray magnetic fields at low frequencies that cause the so-called ``ground loops''.

While a low frequency noise component can be removed by 
using isolation transformers and capacitor-coupled inputs to the ADCs,
that approach is not an option in high-rate environments
where AC-coupled devices would produce unacceptable 
signal distortions and rate-dependent baseline shifts.

Custom electronic circuits developed to address the problem of
low frequency voltage ripples are described in Refs.~\cite{Gar96,Llo99}. 
They are designed to provide a correspondence between a simple 
saw-tooth waveform and an experimental AC-power cycle, digitizing that 
information and passing it to the data stream for an offline noise correction.

Another method that uses one or more ``blackened PMTs'' (bPMTs) 
in parallel with the detector active PMTs is described and compared with 
active AC-noise synchronization circuits in Ref.~\cite{Llo99}. 
The bPMTs are dummy phototubes under high voltage but with no attached detector
modules, whose signals are digitized in exactly the same way as those of 
the active detectors. A drawback/disadvantage of the method is that for 
complicated experimental layouts operating in noisy environments and 
with multiple local grounds prevailing in the areas close to beamlines, 
more than a handful of the bPMTs would be necessary to account properly for 
the correlated noise. Moreover, because bPMTs should be mechanically and 
electrically a part of a detector and should be physically close the
active PMTs, they could be affected by, for example, Cherenkov radiation
caused by relativistic minimum ionizing particles in photocathode glass
windows, destroying the noise correlations.

In the analysis that follows we take advantage of the fact that our apparatus,
the PIBETA detector at PSI, consists of individual detectors that are optically 
isolated from each other. Therefore, one particular physics trigger 
will not excite more than a handful of active detectors and the role of 
a ``blackened PMT'' can be played by different active detector lines in different 
events. We call our procedure ``the second pedestal noise correction''.

\bigskip
\section{Experimental layout}
\medskip

The PIBETA apparatus~\cite{Poc88} is a large acceptance 3$\pi\,$sr 
non-magnetic detector optimized for measurements of electrons and photons
in the energy range of 10--100$\,$MeV. The heart of the detector is
a spherical 240-module pure CsI calorimeter that is supplemented
with a segmented 9-piece active plastic target, a pair of 
conventional cylindrical MWPCs, and a cylindrical 20-counter plastic 
hodoscope for charged particle tracking and identification. 

All parts of the detector are mounted at the center of a 2.5$\,$m$\times$5.6$\,$m 
steel platform that also carries high voltage power supplies, 
a detector cooling system, coiled coaxial delay cables on the
one side, and fast trigger electronics on the opposite side of the platform.
Therefore, the PIBETA detector is a compact self-contained assembly
that can be moved easily from its parking place into a chosen 
experimental area as a single unit and made operational within a day. 
Due to the detector's proximity to elements of the beamline, our
fast electronics are exposed to a significant contaminating electronic
hum. Fig.~\ref{fig:waveform} shows a typical baseline of an analog 
signal that is to be digitized in an ADC device, captured on 
a Tektronix TDS 744 digital oscilloscope.
A snapshot displays ground-loop noise with frequencies of
$\sim$50$\,$Hz and $\sim$300$\,$Hz in a 20$\,\mu$s interval.
Typical peak-to-peak noise amplitude is $<$5$\,$mV.

PIBETA PMT signal outputs are split either directly at the PMT voltage dividers
or at passive custom-made analog signal splitters. One branch of a calorimeter
analog signal is delayed $\sim$380$\,$ns in coaxial cables and then split again
to provide inputs for FASTBUS ADC units and discriminators/TDCs/scalers.
The other branch of the PMT signal is connected to analog summing and discriminator
modules of the fast trigger electronics. The master trigger outputs are 
delayed and subsequently used to provide ADC gates and TDC start/stop signals.   
Metal frames of the PMT high voltage supplies on one end of the platform
and the detector conductive support structure in the center are connected 
with 10$\,$mm/20$\,$mm thick copper cables to fast trigger electronics 
grounding points in order to decrease the noise arising from the ground 
loops. Voltage differences between different parts of the detector are 
measured with a digital voltmeter and after the grounding connections
are put in place are reduced to less than 4$\,$mV.

For digitizing fast PMT pulses we have used a LeCroy ADC model 1882F unit, 
a high-density FASTBUS converter with 96 current-integrating negative 
inputs~\cite{LCR95}. The 1885F unit provided a wide dynamic range 
with 12-bit data for the resolution of 0.05$\,$pC/count and less than
1$\,$\% quantization error. Raw pedestals are 
set at about 200-300 channels and in our particular experimental
environment had the average root-mean-square (rms) widths of $\sim$15-20 
channels.

Fig.~\ref{fig:baseline} presents an analog signal of a single CsI calorimeter 
detector on 2$\,$mV/di\-vi\-sion and 100$\,$ns/di\-vi\-sion scale taken in
6$\,$sec persistence mode. Clearly, with an ADC gate width set to 
100$\,$ns, the ADC unit will integrate variable amounts of the noise current.
A 70$\,$MeV electron or photon will produce a current pulse
with an amplitude of $\sim$0.25$\,$V and a $\sim$30$\,$ns FWHM width.
This current pulse into 50$\,\Omega$ will integrate to a charge of
about 150$\,$pC, placing it at 3/4 of the full scale.  A common noise with an amplitude 
5$\,$mV present on the baseline will integrate in a 100$\,$ns gate to a 10$\,$pC 
charge, or almost 7\% of an actual particle signal. 
As the energies and/or directions of measured particles are deduced from 
ADCs of a centrally hit calorimeter module and on average two nearest 
neighbors hit, the resulting uncertainty in energy (direction) could add up to 20\%.

\section{Random (Pedestal) Trigger}
\medskip
A 190$\times$20$\times$8$\,$mm$^3$ plastic scintillator counter (BC-400) is 
placed horizontally above the electronics racks, about 3$\,$m away from 
the main PIBETA detector. By virtue of its position, the counter is 
shielded from the experimental radiation area by a 5$\,$cm thick lead 
brick wall as well as by a 50$\,$cm wide concrete wall. Operating with 
a high discriminator threshold it is counting only random cosmic muons 
at about 1-2$\,$Hz frequency, having a stable counting rate for both beam-on 
and beam-off periods.

The discriminated signals from that scintillator counter define our random 
trigger. The main PIBETA detector trigger is defined dynamically in 
a programmable logic unit. In practice, the master trigger is a logic OR of 
up to 12 individual triggers. High rate physics triggers in the mix are prescaled, 
with prescaling factors depending on the trigger type and the beam stopping rate.
The random trigger is connected to the trigger mix logic unit
and in production runs it is always included among the enabled triggers.
A single production run is limited to 200,000 events, yielding typically 
$\sim$10$^4$ random triggers per run interspersed among the physics triggers.

We note parenthetically that in an off-line analysis ADC values for all 
detectors in random events are written out to a text file. These data sets
are used in a {\tt GEANT}~\cite{Bru87} detector simulation, enabling us to 
account rigorously for residual ADC noise and accidental event pile-ups 
in the energy spectra.   
\bigskip
\section{Common Noise Groups}
\medskip
In a pre-production calibration run data are first collected in so-called 
``ped\-es\-tal mode'' with the beam off and with only random trigger enabled in
the trigger mix. Raw pedestals are automatically calculated at the end of 
the calibration run. Namely, the pedestal histograms are filled with the random 
trigger events $R_{ni}$ expressed in ADC channel numbers, where a subscript $n$ 
labels a detector and an index $i$ represents an event number. At the end of 
the run these histograms are analyzed for the mean pedestal positions $P_n$. 
The mean pedestal positions and widths are saved in the data acquisition database 
residing completely in shared memory to allow for fast access.
``Calibrated ADC values'' $C_i$ in the subsequent production runs would then be 
differences between raw ADC readings $R_{ni}$ and the pedestal values $P_n$ retrieved 
from the database:
\begin{equation}
C_{ni}=R_{ni}-P_n.
\end{equation}
The analyzer calibration program also calculates the correlation 
coefficients $r_{nm}$ among all possible detector pairs ``$nm$'' that belong 
to the same detector group, e.g., for 240 calorimeter detectors, 40 plastic 
veto hodoscope counters, etc. The expression used for calculating the
correlation coefficient $r_{nm}$ is~\cite{Bev92}:
\begin{equation}
r_{nm}= { { N \sum_{i=1}^N C_{ni} C_{mi} -\sum_{i=1}^N C_{ni} \sum_{i=1}^N C_{mi} } \over
     { \sqrt { N \sum_{i=1}^N (C_{ni})^2 - (\sum_{i=1}^N C_{ni})^2 }
       \sqrt { N \sum_{i=1}^N (C_{mi})^2 - (\sum_{i=1}^N C_{mi})^2 }
     }
   },
\end{equation}
where sums extend over all pedestal events with calibrated ADC values $C_{ni}$ 
less than 20 channels and $N$ is the total number of such events.
The 20 channel limit corresponds to the half-width at tenth-maximum of the
typical noise distribution and discriminated against (very rare) cosmic events 
in our random data set that would smear up our correlation matrix.

The correlation coefficient matrix is saved in a text file that
is analyzed by an offline computer program to identify the detector 
noise groups. The program user can select the maximum number of 
noise groups (default 10), minimum group size (default 8) and the minimum 
correlation coefficient (default 0.75) between the detectors in the same 
noise group. The offline code combines detectors into noise groups on 
a trial-and-error basis with the over-all goal of maximizing the average intergroup
correlation coefficients, while keeping the number of groups as small as 
possible and the minimum number of channels in any group above a given limit.
With a minimum group size of eight detectors the program generated 10 noise
groups for the 240 CsI channels with an average intergroup 
correlation coefficient 0.90 and minimum correlation coefficient 
greater than 0.75.

The noise groups suggested by the program are subsequently hard-coded in 
the analyzer calibration program. An example of six plastic hodoscope (PV) 
noise groups is given below in a section of {\tt C} code:
\begin{verbatim}
/* ADC groups for common noise suppression */

#define MAX_NOISE_GROUP 100
int adc_group[][MAX_NOISE_GROUP] = {

  /* Plastic Veto Detectors */

  {240,241,242,243,244,245,246,247,248,
   249,250,251,252,253,254,255, -1},
  {256,257,258,259, -1},
  {260,261,262,263,264, -1}, 
  {265,266,267,268,269,270,271, -1},
  {272,273,274,275, -1},
  {276,277,278,279, -1},
  { -1 },                          };
\end{verbatim}
where individual PV counters are identified by their signal indices. The value 
``$-1$'' at the end of each line follows the last detector in every noise group. Each 
detector is included in one and only one noise group. Detectors belonging
to the same noise group are usually physically close and have their signal and 
high voltage cables bundled together.

The algorithm for calculating of noise-corrected and calibrated ADC values resides 
in the analyzer program with predefined noise groups. The algorithm loops over 
all ADC channels and for every event:
\begin{enumerate}
\item subtracts the raw pedestal values retrieved from the online database
      from raw ADC variables (``first pedestal correction'');
\item identifies the common noise group that the current detector belongs to;
\item finds the channel with the smallest ADC value in that group;
\item finds all the channels in the group under consideration which are  
      within 15 counts above the minimum of the group and calculates their 
      average ADC value;
\item subtracts that average value (``second pedestal correction'')
      from the ADC values of all other group members.
\end{enumerate}
At the beginning of each run the pedestal histograms are reset and at the end of the run 
they are again analyzed by fitting lineshapes with a Gaussian function. The means 
of the distributions are updated in the database as new pedestal values $P_n$.
\bigskip
\section{Results and Discussion}
\medskip
Fig.~\ref{fig:correlation} shows a scatter plot of ADC values
for two different CsI detectors acquired with a random trigger. 
Uncorrected ADC values are plotted in the top panel showing the correlation 
in an electronic noise: these two detectors are eventually assigned to the same
electronics noise group. Using the secondary pedestal correction procedure
described above, we obtained a scatter-plot of the corrected ADC values
presented in the lower panel of Fig.~\ref{fig:correlation}.

Examples of one-dimensional ADC histograms for two different detector types,
namely a plastic scintillator counter and one active target segment are shown 
in Figs.~\ref{fig:improve}~and~\ref{fig:target}, respectively. Top panels 
represent raw ADC spectra, bottom panels show the noise
suppression in the corrected spectra.

In Fig.~\ref{fig:improve} a peak at channel number 6 is inserted by the fitting 
function in the analyzer program. Its position corresponds to 3 standard
deviations of the corrected pedestal lineshape. In readout of the 
ADC modules all channels with the values below 3$\sigma$ are suppressed by
being considered equal to zero and are not written to tape. This 
zero suppression criterion compresses the size of our data files
by an order of magnitude.

A comparison between the smearing of raw pedestal data and the spectra
obtained after the noise suppression for four different detector types
is given in Table~\ref{tab1}. The table summarizes the maximum
and minimum noise group sizes, rms widths of raw and corrected pedestal 
histograms as well as an important equivalent energy deposition corresponding 
to the corrected pedestal rms values. We see that the energy equivalents of 
the rms widths range from 0.13$\,$MeV in CsI detectors to 0.40$\,$MeV in 
the segmented target. The target pedestals are more difficult to correct 
because of (a) the small number of target segments (9) that form a noise group, 
and (b) the high rate of physics triggers in the target, with up to 1$\,$MHz 
stopping pions. 
\bigskip
\section{Conclusion}
\medskip
We have implemented correction of modular detector ADC readings for the correlated 
common noise using the method of common noise detector groups. The reduction 
in the pedestal width of up to a factor of five is achieved with the corrected
pedestal rms values as narrow as 1.5 ADC channels. This improvement compares 
favorably with the active circuit noise suppression methods~\cite{Gar96,Llo99} 
where improvement factors are $\sim$5-6.

In a representative case of pure CsI calorimeter the procedure reduces 
the correlated noise contribution to the equivalent energy of 0.13$\,$MeV 
per detector module. For the energy range of interest, 10--100$\,$MeV,
this value is smaller than the photoelectron statistics contribution of 
$\sqrt{E/60}\,$MeV~\cite{Frl00} and is comparable with the PMT dark current 
noise that amounts to $\sim$0.1$\,$MeV.
\bigskip
\section{Acknowledgements}
\medskip
This work is supported and made possible by grants from the US National
Science Foundation and the Paul Scherrer Institute.
\vfill\eject

\clearpage

\vspace*{\stretch{1}}
\begin{figure}[!tpb]
\caption{Typical low frequency noise in the electronics shack
next to the $\pi$E1 beam area at the Paul Scherrer Institute. The average amplitude 
of the correlated noise is $<$5$\,$mV and the frequency range 50--500$\,$Hz.}
\label{fig:waveform}
\end{figure}

\begin{figure}[!tpb]
\caption{A typical analog signal baseline for a single CsI PIBETA
detector module. The detector is part of the 240-element PIBETA calorimeter. 
The baseline oscillation is $\sim$3$\,$mV, which is integrated over
a 100$\,$ns ADC gate.}
\label{fig:baseline}
\end{figure}

\begin{figure}[!tpb]
\caption{ADC signal correlation for two CsI detectors 
making the 240-module PIBETA electromagnetic calorimeter. 
The trigger is a random event trigger.}
\label{fig:correlation}
\end{figure}

\begin{figure}[!tpb]
\caption{Improvement in the pedestal width for one plastic veto scintillator
making up the 40-stave PIBETA hodoscope. Dashed line represents 
the raw pedestals, the solid line shows the corrected pedestals.}
\label{fig:improve}
\end{figure}

\begin{figure}[!tpb]
\caption{Improvement in the pedestal width for one channel of
the 9-piece segmented active target after the common noise correction.}
\label{fig:target}
\end{figure}
\vspace*{\stretch{2}}
\clearpage

\bigskip
\begin{table}[ph]
\caption{Comparison of the pedestal noise reduction for the various 
elements of the PIBETA detector. The values represent averages in one 
typical production run with $\sim$5000 pedestal events. The minimum and 
maximum number of detectors in one noise group is given in the second column. 
The energy (angle) equivalent of the corrected pedestal rms is listed in 
the fifth column.}
\bigskip
\label{tab1}
\begin{tabular}{lcccc}
\hline
\multicolumn{1}{c}{Detector}
&\multicolumn{1}{c}{Group Size} 
&\multicolumn{1}{c}{Raw Pedestal} 
&\multicolumn{1}{c}{Corrected Ped.}
&\multicolumn{1}{c}{Corr. Ped. En.}\\
\multicolumn{1}{c}{Type (Number)} 
&\multicolumn{1}{c}{Min/Max}
&\multicolumn{1}{c}{$<$rms$>$ (ch)} 
&\multicolumn{1}{c}{$<$rms$>$ (ch)}
&\multicolumn{1}{c}{(Angle) Equivalent} \\
\hline\hline
CsI Calorimeter Module (240)    & 13/48     & 20.9  & 3.3 & 0.13$\,$MeV \\
Plastic Scintillator Stave (40) &  4/16     & 16.9  & 2.7 & 0.01$\,$MeV \\
Active Plastic Target (9)       &  9/9      & 21.1  & 4.6 & 0.40$\,$MeV \\
MWPC Cathode Strip (192/384)    & 128/384   & 104   & 35  & 0.06$^\circ$ \\
\hline
\end{tabular}
\end{table}
\clearpage

\vspace*{\stretch{1}}
\centerline{\psfig{figure=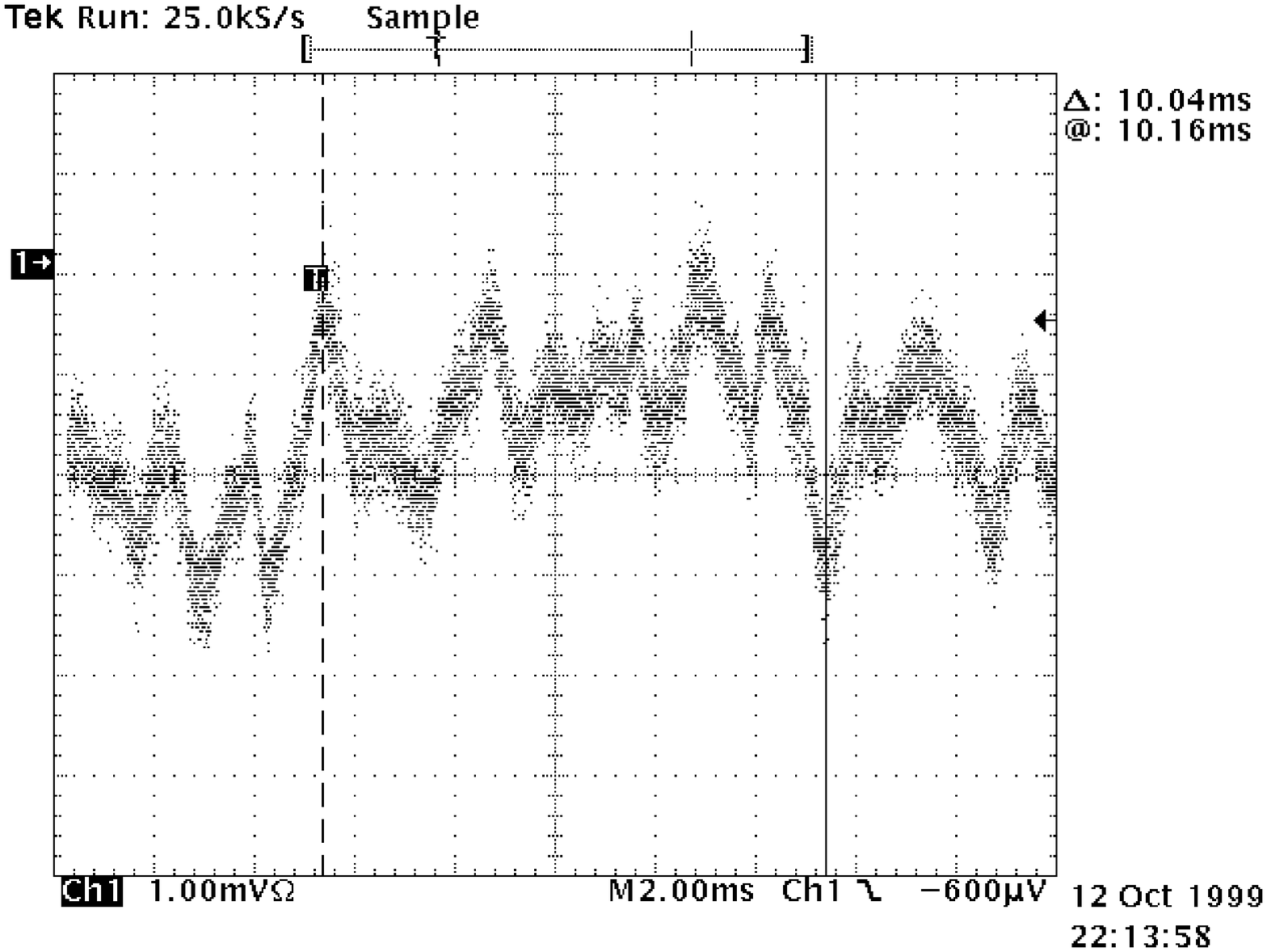,width=16cm}}
\vglue 1cm
\centerline{FIGURE 1}
\vspace*{\stretch{2}}
\clearpage

\vspace*{\stretch{1}}
\centerline{\psfig{figure=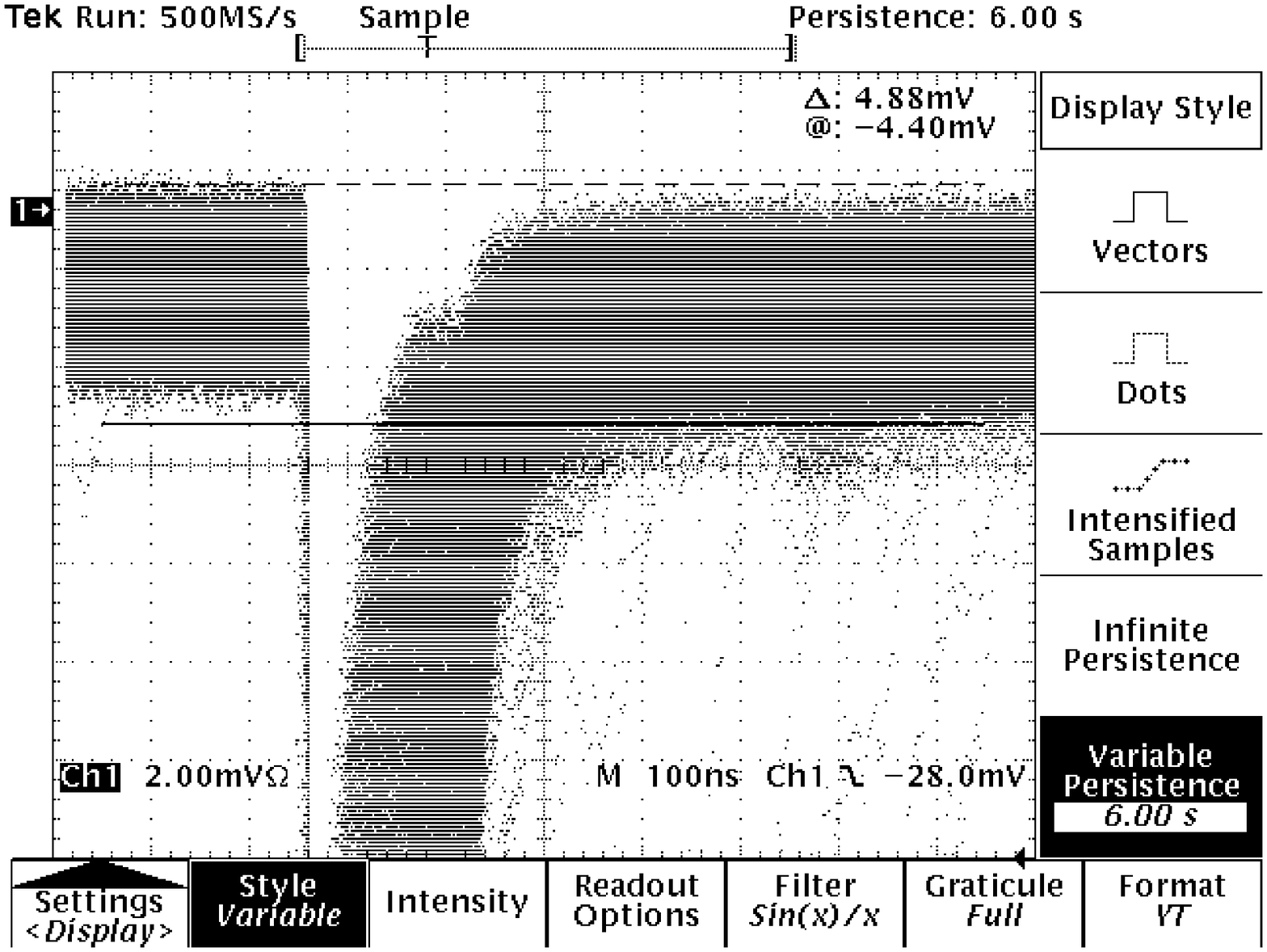,width=16cm}}
\vglue 1cm
\centerline{FIGURE 2}
\vspace*{\stretch{2}}
\clearpage

\vspace*{\stretch{1}}
\centerline{\psfig{figure=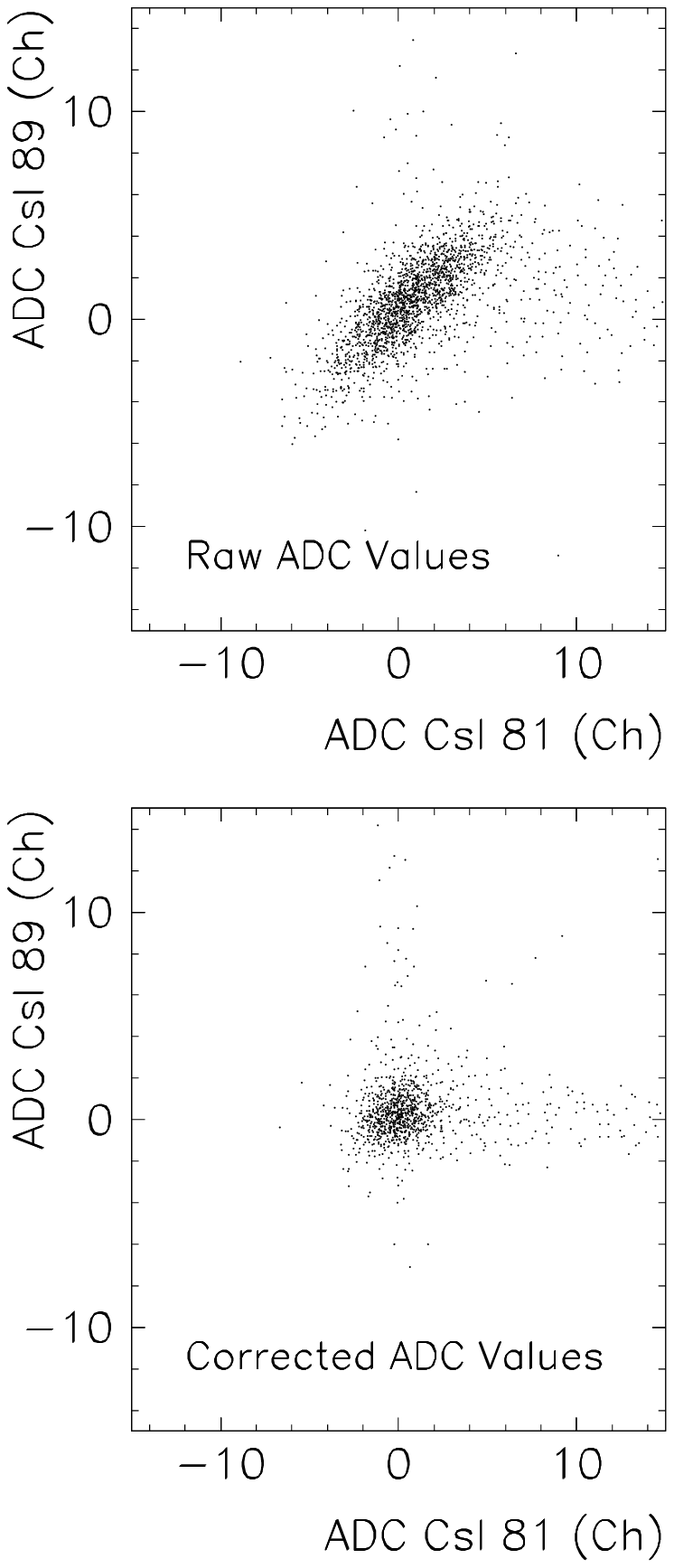,width=11cm}}
\vglue -0.5cm
\centerline{FIGURE 3}
\vspace*{\stretch{2}}
\clearpage

\vspace*{\stretch{1}}
\centerline{\psfig{figure=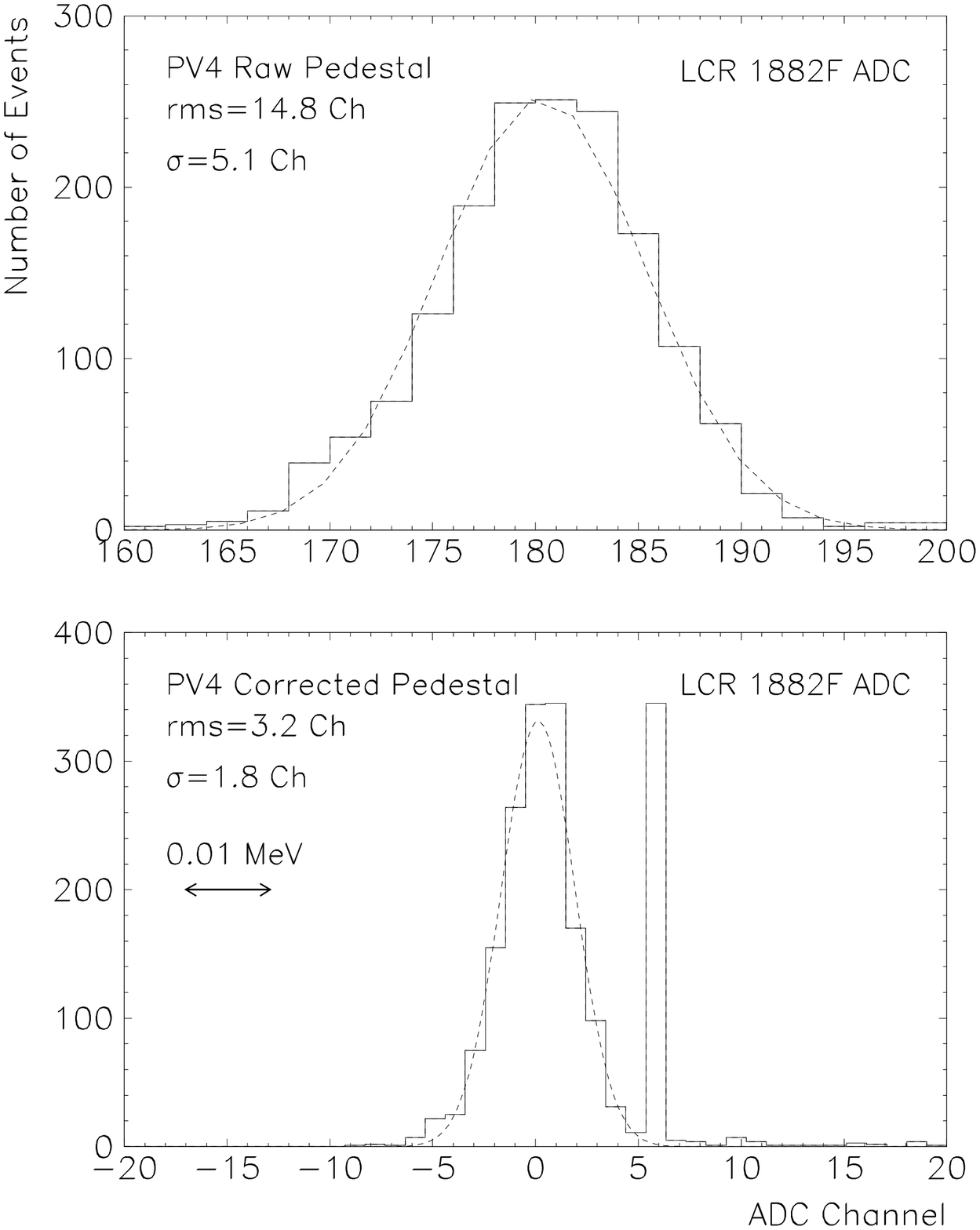,height=22cm}}
\vglue -0.5cm
\centerline{FIGURE 4}
\vspace*{\stretch{2}}
\clearpage

\vspace*{\stretch{1}}
\centerline{\psfig{figure=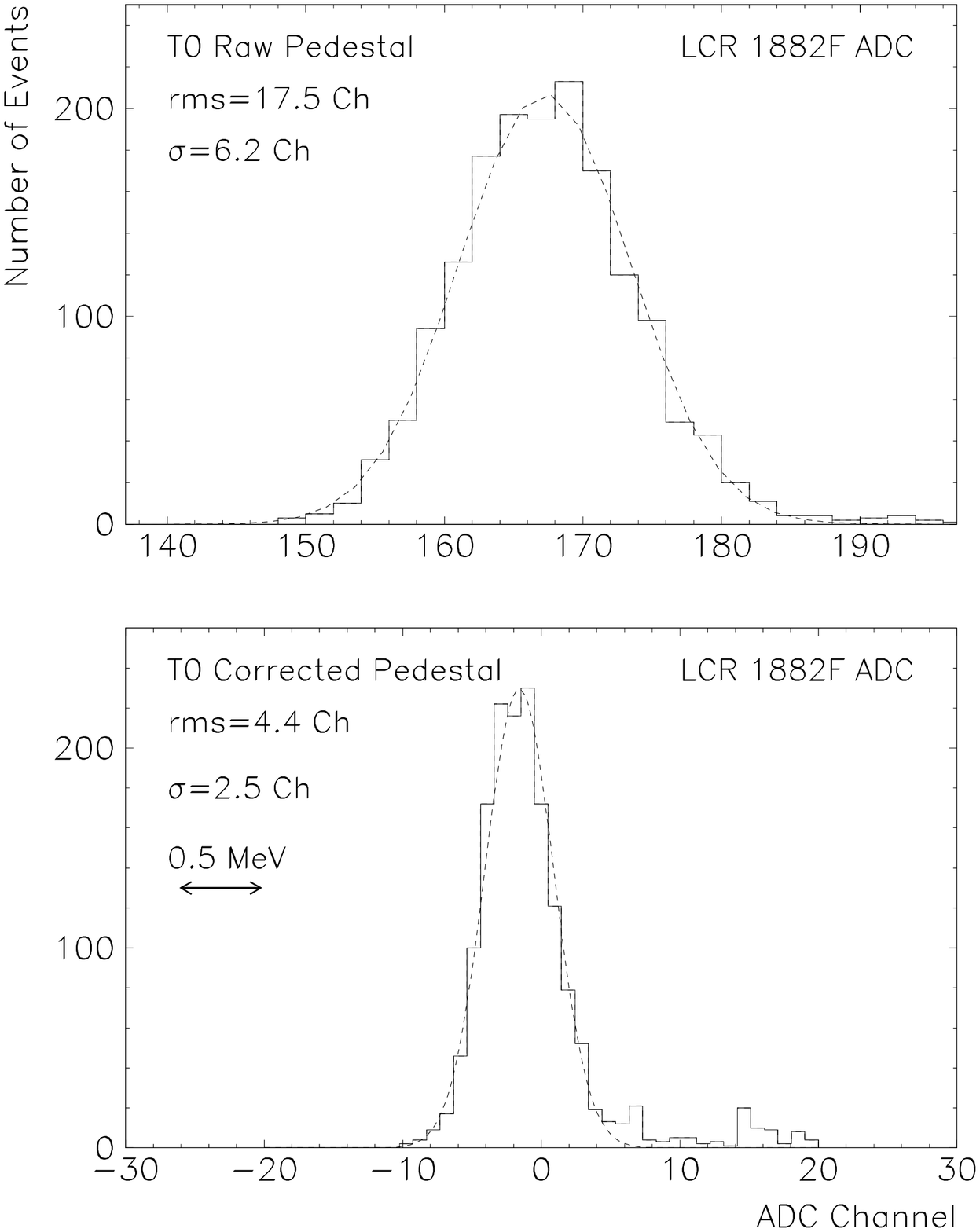,height=22cm}}
\vglue -0.5cm
\centerline{FIGURE 5}
\vspace*{\stretch{2}}
\clearpage

\end{document}